\magnification=\magstep1
\pretolerance 2000
\baselineskip=14pt
\overfullrule=0pt

\catcode`@=11
\def\vfootnote#1{\insert\footins\bgroup\baselineskip=12pt
  \interlinepenalty\interfootnotelinepenalty
  \splittopskip\ht\strutbox 
  \splitmaxdepth\dp\strutbox \floatingpenalty\@MM
  \leftskip\z@skip \rightskip\z@skip \spaceskip\z@skip \xspaceskip\z@skip
  \textindent{#1}\footstrut\futurelet\next\fo@t}
\skip\footins 20pt plus4pt minus4pt
\def\footstrut{\vbox to2\splittopskip{}}
\catcode`@=12
\def\buildurel#1\under#2{\mathrel{\mathop{\kern0pt #2}\limits_{#1}}}
\hfill November 1, 1994
\vglue 24pt plus 12pt minus 12pt
\centerline{\bf MICROSCOPIC REVERSIBILITY AND MACROSCOPIC BEHAVIOR:}
\centerline{\bf PHYSICAL EXPLANATIONS AND MATHEMATICAL DERIVATIONS}
\medskip
\centerline{Joel L. Lebowitz}

\centerline{Departments of Mathematics and Physics}

\centerline{Rutgers University, New Brunswick, NJ 08903}
\bigskip
\bigskip
\bigskip

\bigskip
\bigskip
\noindent {\bf Abstract}
\medskip

The observed general time-asymmetric behavior of 
macroscopic systems---embodied in the second law of thermodynamics---arises
naturally {}from time-symmetric microscopic laws due to the great disparity
between
macro and micro-scales.  More specific features of macroscopic evolution
depend on the nature of the microscopic dynamics.  In particular, short
range interactions with good mixing properties lead, for simple systems,
to the quantitative description of such
evolutions by means of autonomous hydrodynamic equations, e.g.\ the
diffusion equation.

These deterministic time-asymmetric equations accurately describe the
observed behavior 
of {\it individual}
macro systems.  Derivations using ensembles (or
probability distributions) must therefore, to be relevant, 
hold for almost all members of the ensemble, i.e.\ occur with
probability close to one.  Equating observed irreversible macroscopic behavior
with the time evolution of ensembles describing systems having only a few
degrees of freedom, where no such typicality holds, is misguided and
misleading. \bigskip \bigskip
\vfill \eject

\line{\hfill\vbox{\hsize3.25truein\baselineskip10pt
{\it ``The equations of motion in abstract dynamics are
perfectly reversible; any solution of these equations remains valid when
the time variable $t$ is replaced by $-t$.  Physical processes, on the
other hand, are irreversible:  for example, the friction of solids,
conduction of heat, and diffusion.  Nevertheless, the principle of
dissipation of energy is compatible with a molecular theory in which each
particle is subject to the laws of abstract dynamics.''}}}
\hfill W. Thomson, (1874)[1]

\noindent  {\bf Introduction}

Given the success of the statistical approach, pioneered by James
Maxwell, William Thomson (later Lord Kelvin) and made quantitative by
Ludwig Boltzmann, in both explaining and predicting the observed
behavior of macroscopic systems on the basis of their reversible
microscopic dynamics, it is quite surprising that there is still so
much confusion about the ``problem of irreversibility''.  I attribute
this to the fact that the originality of these ideas made them
difficult to grasp.  When put into high relief by Boltzmann's precise
and elegant form of his famous kinetic equation and H-theorem they
became ready targets for attack.  The confusion created by these
misunderstandings and by the resulting ``controversies'' between
Boltzmann and some of his contemporaries, particularly Ernst Zermelo,
has been perpetuated by various authors who either did not understand
or did not explain adequately the completely satisfactory resolution
of these questions by Boltzmann's responses and later writings.  There
is really no excuse for this, considering the clarity of the latter.
In Erwin Schr\"odinger's words, ``Boltzmann's ideas really give an
understanding" of the origin of macroscopic behavior. All claims of
inconsistencies (known to me) are in my opinion wrong and I see no
need to search for alternate explanations of such behavior---at least
on the non-relativistic classical level .  I highly recommend some of
Boltzmann's works [2], as well as the beautiful 1874 paper of Thomson
[1] and the more contemporary references [3-7], for further reading on
this subject; see also [8] for more details on the topics discussed
here.

Boltzmann's statistical theory of nonequilibrium (time-asymmetric,
irreversible) behavior associates {\it to each microscopic state of a
macroscopic system}, be it gas, fluid or solid, a number $S_B$: the
``Boltzmann entropy" of that state [4].  This entropy agrees (up to
terms negligible in the size of the system) with the macroscopic
thermodynamic entropy of Clausius {\it when} the system is in
equilibrium.  It also coincides {\it then} with the Gibbs entropy
$S_G$, which is defined not for individual microstates but for
statistical ensembles or probability distributions (in a way to be
described later).  The agreement extends to systems in local
equilibrium.  However, unlike $S_G$, which does not change in time
even for ensembles describing (isolated) systems not in equilibrium,
e.g.\ fluids evolving according to hydrodynamic equations, $S_B$
typically increases in a way which {\it explains} and describes
qualitatively the evolution towards equilibrium of macroscopic
systems.

This behavior of $S_B$ is due to the separation between microscopic
and macroscopic scales, i.e.\ the very large number of degrees of
freedom involved in the specification of macroscopic properties.  It
is this separation of scales which enables us to make definite
predictions about the evolution of a {\it typical individual
realization} of a macroscopic system, where, after all, we actually
observe irreversible behavior.  As put succinctly by Maxwell [9] ``the
second law is drawn {}from our experience of bodies consisting of an
immense number of molecules. \dots it is continually being violated,
\dots, in any sufficiently small group of molecules \dots .  As the
number \dots is increased \dots the probability of a measurable
variation \dots may be regarded as practically an impossibility''.
The various ensembles commonly used in statistical mechanics are to be
thought of as nothing more than mathematical tools for describing
behavior which is practically the same for ``almost all'' individual
macroscopic systems in the ensemble.  While these tools can be very
useful and some theorems that are proven about them are very beautiful
they must not be confused with the real thing going on in a single
system.  To do that is to commit the scientific equivalent of
idolatry, i.e.\ substituting representative images for reality.
Moreover, the time-asymmetric behavior manifested in a single typical
evolution of a macroscopic system distinguishes macroscopic
irreversibility {}from the mixing type of evolution of ensembles which
are caused by the {\it chaotic} behavior of systems with but a few
degrees of freedom, e.g.\ two hard spheres in a box. To call the
latter irreversible is, therefore, confusing.

The essential qualitative features of macroscopic behavior can be
understood on the basis of the incompressible flow in phase space
given by Hamilton's equations.  They are not dependent on assumptions,
such as positivity of Lyapunov exponents, ergodicity, mixing or
``equal a priori probabilities,'' being {\it strictly} satisfied.
Such properties are however important for the quantitative description
of the macroscopic evolution which is given, in many cases, by
time-asymmetric autonomous equations of hydrodynamic type.  These can
be derived (rigorously, in some cases) {}from reversible microscopic
dynamics by suitably scaling macro and micro units of space and time
and then taking limits in which the ratio of macroscopic to
microscopic scales goes to infinity [10].  (These limits express in a
mathematical form the physics arising {}from the very large ratio of
macroscopic to microscopic scales.)  Using the law of large numbers
then shows that these equations describe the behavior of almost all
individual systems in the ensemble, not just that of ensemble
averages, i.e.\ the dispersion goes to zero in the scaling limit.
Such descriptions also hold, to a high accuracy, when the macro/micro
ratio is finite but very large. They are however clearly impossible
when the system contains only a few particles.

The existence and form of such hydrodynamic equations depends on the
nature of the microscopic dynamics.  In particular, instabilities of
trajectories induced by chaotic microscopic dynamics play an important
role in determining many features of macroscopic evolution.  A simple
example in which this can be worked out in detail is provided by the
Lorentz gas.  This consists of a macroscopic number of non-interacting
particles moving among a periodic array of fixed convex scatterers
arranged in the plane so that there is a maximum distance a particle
can travel between collisions. The chaotic nature of the microscopic
dynamics, which leads to an approximately isotropic local distribution
of velocities, is directly responsible for the existence of a simple
autonomous deterministic description, via a diffusion equation, for
{\it typical} macroscopic particle density profiles of this system
[10].  Another example is the description via the Boltzmann equation
of the density in the six dimensional position and velocity space of a
macroscopic dilute system of hard spheres [7], [10].  I use these
examples, despite their highly idealized nature, because here all the
mathematical $i$'s have been dotted.  They thus show {\it ipso facto},
in a way that should convince even (as Mark Kac put it) an
``unreasonable" person, not only that there is no conflict between
reversible microscopic and irreversible macroscopic behavior but also
that, {\it for essentially all initial microscopic states consistent
with a given nonequilibrium macroscopic state}, the latter follows
{}from the former---in complete accord with Boltzmann's ideas.

Boltzmann's analysis was of course done in terms of classical
Newtonian mechanics and I shall use the same framework for this
article.  The situation is in many ways similar in quantum mechanics
where reversible incompressible flow in phase space is replaced by
unitary evolution in Hilbert space.  In particular I do not believe
that quantum measurement is a {\it new} source of irreversibility.
Such assertions in effect ``put the cart before the horse''.  Real
measurements on quantum systems are time-asymmetric because they
involve, of necessity, systems with very large number of degrees of
freedom whose irreversibility can be understood using natural
extensions of classical ideas [11], [13].

There are however also some genuinely new features in quantum
mechanics relevant to our problem. First, to follow the classical
analogy directly one would have to associate a macroscopic state to an
arbitrary wave function of the system, which is impossible as is clear
{}from the Schr{\"o}dinger cat paradigm [12] (or paradox).  Second,
quantum correlations between separated systems arising {}from wave
function entanglements lead to the impossibility, in general, of
assigning a wave function to a subsystem ${\cal S}_1$ of a system
$\cal S$ in a definite state $\psi$ even at a time when there is no
direct interaction between ${\cal S}_1$ and the rest of $\cal S$, and
this makes the idealization of an isolated system much more
problematical in quantum mechanics than in classical theory.  These
features of quantum mechanics require careful analysis to see how they
affect the irreversibility observed in the real world.  An in depth
discussion is not only beyond the scope of this article but would also
require some new ideas and quite a bit of work which is yet to be
done.  I refer the reader to references [11--14] for a discussion of
some of these questions {}from many points of view.

I will also, in this article, completely ignore relativity, special or
general. The phenomenon we wish to explain, namely the time-asymmetric
behavior of spatially localized macroscopic objects, has certainly
many aspects which are the same in the relativistic (real) universe as
in a (model) non-relativistic one. This means of course that I will
not even attempt to touch the deep conceptual questions regarding the
nature of space and time itself which have been much discussed
recently in connection with reversibility in black hole radiation and
evaporation [14].  These are beyond my competence and indeed it seems
that their resolution may require new concepts which only time will
bring.  I will instead focus on the problem of the origin of
macroscopic irreversibility in the simplest idealized classical
context.  The Maxwell-Thomson-Boltzmann resolution of this problem in
these models does, in my opinion, carry over essentially unchanged to
real systems.

\bigskip
\noindent    {\bf The Problem of Macroscopic Irreversibility}

Consider a macroscopic system evolving in time, as exemplified by the
schematic snapshots of a binary system, say two different colored
inks, in the four frames in Figure 1.  The different frames in this
figure represent pictorially the two local concentrations of the
components at different times. {\it Suppose} we know that the system
was isolated during the whole time of picture taking and we are asked
to identify the time order in which the snapshots were taken.

The obvious answer, based on experience, is that time increases {}from
1a to 1d---any other order is clearly absurd.  Now it would be very
simple and nice if this answer could be shown to follow directly
{}from the microscopic laws of nature.  But this is not the case, for
the microscopic laws, as we know them, tell a different story: \ if
the sequence going {}from left to right is permitted by the
microscopic laws, so is the one going {}from right to left.

This is most easily seen in classical mechanics where the complete
microscopic state of an isolated classical system of $N$ particles is
represented by a point $X=({\bf r}_1, {\bf v}_1, {\bf r}_2, {\bf v}_2,
\dots, \ {\bf r}_N, {\bf v}_N)$ in its phase space $\Gamma$, ${\bf
r}_i$ and ${\bf v}_i$ being the position and velocity of the $i$th
particle.  Now a snapshot in Fig.\ 1 clearly does not specify
completely the microstate $X$ of the system; rather each picture
specifies a coarse grained description of $X$, which we denote by
$M(X)$, the macrostate corresponding to $X$.  For example, if we
imagine that the (one liter) box in Fig.\ 1 is divided into a billion
little cubes, then the macrostate $M$ could simply specify (within
some tolerance) the fraction of particles of each type in every cube
$j$, $j=1$, $\dots$, $10^9$.  To each macrostate $M$ there corresponds
a very large set of microstates making up a region $\Gamma_M$ in the
phase space $\Gamma$.  In order to specify properly the region
$\Gamma_M$ we need to know also the total energy $E$, and any other
{\it macroscopically relevant}, e.g.\ additive, constants of the
motion (also within some tolerance). While this specification of the
macroscopic state clearly contains some arbitrariness, this need not
concern us unduly here.  All the qualitative statements we are going
to make about the time evolution of macrostates $M$ are sensibly
independent of its precise definition as long as there is a large
separation between the macro and microscales.

Let us consider now the time evolution of microstates which underlies
that of the macrostates $M(X)$.  They are governed by Hamiltonian
dynamics which connects a microstate $X(t_0)$ at some time $t_0$, to
the microstate $X(t)$ at any other time $t$.  Let $X(t_0)$ and $X(t_0
+ \tau)$, {}~ $\tau > 0$, be two such microstates.  Reversing
(physically or mathematically) all velocities at time $t_0 + \tau$, we
obtain a new microstate.  If we now follow the evolution for another
interval $\tau$ we arrive at a microstate at time $t_0 + 2 \tau$ which
is just the state $X(t_0)$ with all velocities reversed.  We shall
call $RX$ the microstate obtained {}from $X$ by velocity reversal, $RX
= ({\bf r}_1, -{\bf v}_1, {\bf r}_2, - {\bf v}_2, \dots , {\bf r}_N,
-{\bf v}_N).$

Returning now to the snapshots shown in the figure it is clear that
they would remain unchanged if we reversed the velocities of all the
particles; hence if $X$ belongs to $\Gamma_M$ then also $RX$ belongs
to $\Gamma_M$.  Now we see the problem with our definite assignment of
a time order to the snapshots in the figure: that a macrostate $M_1$
at time $t_1$ evolves to another macrostate $M_2$ at time $t_2 = t_1 +
\tau$, $\tau > 0$, means that there is a microstate $X$ in
$\Gamma_{M_1}$ which gives rise to a microstate $Y$ at $t_2$ with $Y$
in $\Gamma_{M_2}$.  But then $RY$ is also in $\Gamma_{M_2}$ and
following the evolution of $RY$ for a time $\tau$ would produce the
state $RX$ which would then be in $\Gamma_{M_1}$.  Hence the snapshots
depicting $M_a$, $M_b$, $ M_c$ and $M_d$ in Fig. 1 could, as far as
the laws of mechanics (which we take here to be the laws of nature)
go, correspond to a sequence of times going in either direction.

It is thus clear that our judgement of the time order in Fig. 1 is not
based on the dynamical laws of evolution alone; these permit either order.
Rather it is based on experience: \ one direction is common and easily
arranged, the other is never seen.  {\it But why should this be so}?
\bigskip

\noindent {\bf Boltzmann's Answer}
\medskip
The above question was first raised and the answer developed by
theoretical physicists in the second half of the nineteenth century
when the applicability of the laws of mechanics to thermal phenomena
was established by the experiments of Joule and others.  The key
people were Maxwell, Thomson and Boltzmann.  As already mentioned I
find the 1874 article by Thomson an absolutely beautiful exposition
containing the full qualitative answer to this problem.  This paper
is, as far as I know, never referred to by Boltzmann or by latter
writers on the subject.  It would or should have cleared up many a
misunderstanding.  I can only hope (but do not really expect) that my
article will do better.  Still I will try my best to say it again in
more modern (but less beautiful) language.  The answer can be
summarized by a quote {}from Gibbs which appears (in English) on the
flyleaf of Boltzmann's Lectures on Kinetic Theory, Vol. 2, [15] (in
German): ``In other words, the impossibility of an uncompensated
decrease of entropy seems to be reduced to improbability [16].''

This statistical theory can be best understood by associating to each
macroscopic state $M$ and {\it thus to each phase point $X$ giving
rise to $M$}, a ``Boltzmann entropy'', defined as
$$S_B (M) = k \log \ |\Gamma_M|, \eqno(1)$$
\noindent where $k$ is Boltzmann's constant and 
$|\Gamma_M|$ is the phase space volume associated with the macrostate $M$,
i.e.\ $|\Gamma_M|$ is the integral of the time invariant Liouville volume
element ($\prod\limits^N_{i=1}d^3{\bf r}_i\; d^3{\bf v}_i$) over
$\Gamma_M$. ($S_B$ is defined up to additive constants, see [4].)

Boltzmann's stroke of genius was to make a direct connection
between this microscopically defined function $S_B(M(X))$ 
and the thermodynamic entropy of Clausius, $S_{eq}$, which
is a macroscopically defined, operationally measurable (up to additive
constants), extensive property of macroscopic systems in {\it equilibrium}.
For a system in equilibrium having a given energy $E$ (within some
tolerance) volume $V$ and particle number $N$, Boltzmann showed that

$$S_{eq} (E,V,N) = N s_{eq} (e,n) \simeq S_B(M_{eq}), \quad e = E/N,\; n =
N/V,\eqno(2)$$

\noindent where $M_{eq} (E,V,N)$ is the equilibrium macrostate
(corresponding to $M_d$ in Fig.\ 1).  
By the symbol 
$\simeq$ we mean that for large $N$, such that the
system is really macroscopic, the equality holds up to terms negligible 
when both sides of equation (2) are divided by $N$ and the additive
constant is suitably fixed.  It is important that the
cells used to define $M_{eq}$ contain many particles, i.e.\ that the
macroscale be very
large compared with the microscale.

Having made this identification it is natural to use Equation (1) to
also define (macroscopic) entropy for systems not entirely in
equilibrium and thus identify increases in such entropy with increases
in the volume of the phase space region $\Gamma_{M(X)}$.  This
identification explains in a natural way the observation, embodied in
the second law of thermodynamics, that when a constraint is lifted
{}from an isolated macroscopic system, it evolves toward a state with
greater entropy.  To see how the explanation works, imagine that there
was initially a wall dividing the box in Fig.\ 1 which is removed at
time $t_a$.  The phase space volume available to the system without
the wall is fantastically enlarged: If the system in fig.\ 1 contains
1 mole of fluid in a 1-liter container the volume ratio of the
unconstrained region to the constrained one is of order $2^N$ or
$10^{10^{20}}$, roughly the ratio $|\Gamma_{M_d}| / |\Gamma_{M_a}|$.
We can then expect that when the constraint is removed the dynamical
motion of the phase point $X$ will with very high ``probability'' move
into the newly available regions of phase space, for which
$|\Gamma_{M}|$ is large.  This may be expected to continue until
$X(t)$ reaches $\Gamma_{M_{eq}}$ corresponding to the system now being
in its unconstrained equilibrium state.  After that time we can expect
to see only small fluctuations {}from macroscopic
equilibrium---typical fluctuations being of order of the square root
of the number of particles involved.  It should be noted here that an
important ingredient in the whole analysis is the constancy in time of
the Liouville volume of sets in the phase space $\Gamma$.  Without
this invariance the connection between phase space volume and
probability would be impossible or at least very problematic.

Of course, if our isolated system remains isolated 
forever, Poincar{\'e}'s Recurrence 
Theorem tells us that the system phase point $X(t)$ would have to come
back very close to its initial value $X(t_a)$, and do so again and again.
But these Poincar{\'e} recurrence times are so enormous (more or less
comparable to the ratio of $|\Gamma_{M_d}|$ to $|\Gamma_{M_a}|$) that when
Zermelo brought up this objection to Boltzmann's explanation of the
second law, Boltzmann's response [17] was as follows:  
``Poincar{\'e}'s theorem,
which Zermelo explains at the beginning of his paper, is clearly correct,
but his application of it to the theory of heat is not. \dots Thus when
\dots 
Zermelo concludes, {}from the theoretical fact that the initial states
in a gas must recur---without having calculated how long a time this
will take---that the hypotheses of gas theory must be rejected or else
fundamentally changed, he is just like a dice player who has
calculated that the probability of a sequence of 1000 one's is not
zero, and then concludes that his dice must be loaded since he has not
yet observed such a sequence!''\footnote {$^{a)}$}{It is remarkable
that in the same paper Boltzmann also wrote ``likewise, it is observed
that very small particles in a gas execute motions which result {}from
the fact that the pressure on the surface of the particles may
fluctuate''.  This shows that Boltzmann completely understood the
cause of Brownian motion ten years before Einstein's seminal papers on
the subject.  Surprisingly he never used this phenomenon in his
arguments with Ostwald and Mach about the reality of atoms.}

Thus not only did Boltzmann's great insights give a microscopic
interpretation of the mysterious thermodynamic entropy of Clausius; they
also gave a natural generalization of entropy to nonequilibrium macrostates
$M$, and with it an explanation of the second law of thermodynamics---the
formal expression of the time-asymmetric evolution of macroscopic states
occurring in nature.

\bigskip

\noindent{\bf The Use of Probability}
\medskip

Boltzmann's ideas are, as Ruelle [6] says, at the same time simple and
rather subtle.  They introduce into the ``laws of nature'' notions of
probability, which, certainly at that time, were quite alien to the
scientific outlook.  Physical laws were supposed to hold without any
exceptions, not just almost always and indeed no exceptions were (or are)
known to the second law; nor would we expect any, as Richard Feynman [3]
rather conservatively says, ``in a million years''.  The reason for this,
as recognized by Maxwell, Thomson and Boltzmann, is that, for a macroscopic
system, the fraction of microstates for which the evolution leads to
macrostates with larger $S_B$ is so close to one (in terms of their
Liouville volume) that such behavior is exactly what should be seen to
``always'' happen.  As put by Boltzmann [17], ``Maxwell's law of the
distribution of velocities among gas molecules is by no means a theorem of
ordinary mechanics which can be proved {}from the equations of motion alone;
on the contrary, it can only be proved that it has very high probability,
and that for a large number of molecules all other states have by
comparison such a small probability that for practical purposes they can be
ignored.''  In present day mathematical language we say that such behavior is
{\it typical}, by which we mean that the set of microstates $X$ in
$\Gamma_{M_{a}}$ for which it occurs have a volume fraction which goes to $1$
as $N$ increases.  Thus in Fig.\ 1 the sequence going {}from left to right is
typical for a phase point in $\Gamma_{M_a}$ while the  one going {}from right
to left has ``probability'' approaching zero with respect to a uniform
distribution in $\Gamma_{M_d}$, for $N$ tending towards infinity.

Note that Boltzmann's argument does not really require the assumption
 that over very long periods of time the macroscopic system should be
 found in different regions $\Gamma_M$, i.e.\ in different macroscopic
 states $M$, for fractions of time {\it exactly} equal to the ratio of
 $|\Gamma_M|$ to the total phase space volume specified by its energy.
 Such behavior, which can be considered as a mild form of Boltzmann's
 ergodic hypothesis, mild because it is only applied to those regions
 of the phase space representing macrostates $\Gamma_M$, seems very
 plausible in the absence of constants of the motion which decompose
 the energy surface into regions with different macroscopic states.
 It appears even more reasonable when we take into account the lack of
 perfect isolation in practice which will be discussed later.  Its
 implication for ``small fluctuations'' {}from equilibrium is
 certainly consistent with observations.  (The stronger form of the
 ergodic hypothesis also seems like a natural assumption for
 macroscopic systems.  It gives a simple derivation for many
 equilibrium properties of macro systems.)

\bigskip
\noindent {\bf Initial Conditions}

Once we accept the statistical explanation of why macroscopic systems
evolve in a manner that makes $S_B$ increase with time, there remains
the nagging problem (of which Boltzmann was well aware) of what we
mean by ``with time''.  Since the microscopic dynamical laws are
symmetric, the two directions of the time variable are {\it a priori}
equivalent and thus must remain so {\it a posteriori} [18].  In
particular if a system with a nonuniform macroscopic density profile,
such as $M_b$, at time $t_b$ in Fig.\ 1 had a microstate that is
typical for $\Gamma_{M_b}$, then almost surely its macrostate at both
times $t_b + \tau$ and $t_b - \tau$ will be like $M_c$.  This is
inevitable: Since the phase space region $\Gamma_{M_b}$ corresponding
to $M_b$ at some time $t_b$ is invariant under the transformation $X
\to RX$, it must make the same prediction for $t_b - \tau$ as for $t_b
+ \tau$.  Yet experience shows that the assumption of typicality at
time $t_b$ will give the correct behavior only for times $t > t_b$ and
not for times $t < t_b$.  In particular, given just $M_b$ and $M_c$,
we have no hesitation in ordering $M_b$ before $M_c$.

If we think further about our ordering of $M_b$ and $M_c$, we realize
that it seems to derive {}from our assumption that $M_b$ is itself so
unlikely that it must have evolved {}from an initial state of even
lower entropy like $M_a$.  {}From an initial microstate typical of the
macrostate $M_a$, which can be readily created by an experimentalist,
we get monotonic behavior of $S_B$ with the time ordering $M_a$,
$M_b$, $M_c$ and $M_d$.  If, by contrast, the system in Fig.\ 1 had
been completely isolated for a very long time compared with its
hydrodynamic relaxation time, then we would expect to always find it
in its equilibrium state $M_d$ (with possibly some small fluctuations
around it).  Presented instead with the four pictures, we would (in
this very, very unlikely case) have no basis for assigning an order to
them; microscopic reversibility assures that fluctuations {}from
equilibrium are typically symmetric about times at which there is a
local minimum of $S_B$.  In the absence of any knowledge about the
history of the system before and after the sequence of snapshots
presented in Fig.\ 1, we use our experience to conclude that the
low-entropy state $M_a$ must have been an initial prepared state.  In
the words of Roger Penrose [5]: ``The time-asymmetry comes merely
{}from the fact that the system has been {\it started off} in a very
special (i.e.\ low-entropy) state, and having so started the system,
we have watched it evolve in the {\it future} direction''.

The point is that a microstate corresponding to $M_b$ (at time $t_b$)
which comes {}from $M_a$ (at time $t_a$) must be {\it atypical} in
some respects of points in $\Gamma_{M_b}$.  This is so because, by
Liouville's theorem, the set $\Gamma_{ab}$ of all such phase points
has a volume $|\Gamma_{ab}| \leq |\Gamma_{M_a}|$ that is {\it very
much smaller} than $|\Gamma_{M_b}|$.  This need not however prevent
the overwhelming majority of points in $\Gamma_{ab}$ (with respect to
Liouville measure on $\Gamma_{ab}$ which is the same as Liouville
measure on $\Gamma_a$) {}from having future macrostates like those
typical of $\Gamma_b$---while still being very special and
unrepresentative of $\Gamma_{M_b}$ as far as their past macrostates
are concerned.  This sort of behavior is what is explicitly proven by
Lanford in his derivation of the Boltzmann equation [7], and is
implicit in all derivation of hydrodynamic equations [10]; see also
[19].  To see intuitively the origin of such behavior we note that for
systems with realistic interactions the domain $\Gamma_{ab}$ will be
so convoluted that it will be ``essentially dense'' in $\Gamma_b$, so
that any slight thickening of it will cover all of $\Gamma_{M_b}$. It
is therefore not unreasonable that their future behavior, as far as
macrostates go, will be unaffected by their past history.

(This can be worked out completely for a model macroscopic system in
which the (large) $N$ noninteracting atoms are each specified not by
$({\bf r, v})$ but by ${\bf
\sigma} = (\dots, \sigma_{-2}, \sigma_{-1}; \sigma_0,
\sigma_1,\dots)$, a doubly 
infinite sequence of zeros and ones (equivalently a point in the unit
square).  Their discrete time dynamics is that of a shift to the left
$(T{\bf \sigma})_i = \sigma_{i+1}$ (equivalently the baker's
transformation).  If we define ``velocity reversal'' by $(R{\bf
\sigma})_i = \sigma_{-i-1}$ and the macrostate $M({\bf \sigma})$ by
the $k$ values, $M({\bf \sigma}) = (\sigma_0+\sigma_{-1},
\sigma_1+\sigma_{-2},
\sigma_2+\sigma_{-3}, \dots , \sigma_{k-1}+\sigma_{-k})$ then a little
thought shows that the future behavior of typical points in
$\Gamma_{M_{ab}}$ is indeed 
as described above.)

\bigskip
\noindent {\bf Origin of Low-Entropy States}

The creation of low-entropy initial states poses no problem in
laboratory situations such as the one depicted in Fig.\ 1.  Laboratory
systems are prepared in states of low Boltzmann entropy by
experimentalists who are themselves in low-entropy states.  Like other
living beings, they are born in such states and maintained there by
eating nutritious low-entropy foods, which in turn are produced by
plants using low-entropy radiation coming {}from the Sun.  That was
already clear to Boltzmann as may be seen {}from the following quote
[20]: ``The general struggle for existence of living beings is
therefore not a fight for the elements---the elements of all organisms
are available in abundance in air, water, and soil---nor for energy,
which is plentiful in the form of heat, unfortunately untransformably,
in every body.  Rather, it is a struggle for entropy that becomes
available through the flow of energy {}from the hot Sun to the cold
Earth.  To make the fullest use of this energy, the plants spread out
the immeasurable areas of their leaves and harness the Sun's energy by
a process as yet unexplored, before it sinks down to the temperature
level of our Earth, to drive the chemical syntheses of which one has
no inkling as yet in our laboratories.  The products of this chemical
kitchen are the object of the struggles in the animal world''.

Note that while these experimentalists have evolved, thanks to this
source of low entropy energy, into beings able to prepare systems in
particular macrostates with low values of $S_B(M)$, like our state
$M_a$, the total entropy $S_B$, including the entropy of the
experimentalists and that of their environment, must always increase:
There are no Maxwell demons.  The low entropy of the solar system is
also manifested in events in which there is no human
participation---so that, for example, if instead of Fig.\ 1 we are
given snapshots of the Shoemaker-Levy comet and Jupiter before and
after their collision, then the time direction is again obvious.

We must then ask what is the origin of this low entropy state of the
solar system.  In trying to answer this question we are led more or
less inevitably to cosmological considerations of an initial ''state
of the universe'' having a very small Boltzmann entropy.  To again
quote Boltzmann [10]: ``That in nature the transition {}from a
probable to an improbable state does not take place as often as the
converse, can be explained by assuming a very improbable initial state
of the entire universe surrounding us.  This is a reasonable
assumption to make, since it enables us to explain the facts of
experience, and one should not expect to be able to deduce it {}from
anything more fundamental''.  That is, the universe is pictured as
having been ``created'' in an initial microstate $X$ typical of some
macrostate $M_0$ for which $|\Gamma_{M_0}|$ is a very small fraction
of the ``total available'' phase space volume.  In Boltzmann's time
there was no physical theory of what such an initial state might be
and Boltzmann toyed with the idea that it was just a very large, very
improbable, fluctuation in an eternal universe which spends most of
its time in an equilibrium state.  Richard Feynman argues convincingly
against such a view [3].

In the current big bang scenario it is reasonable, as Roger Penrose
does in [5], to take as initial state the state of the universe just
after the big bang.  Its macrostate would then be one in which the
energy density is approximately spatially uniform.  Penrose estimates
that if $M_f$ is the macrostate of the final ``Big Crunch'', having a
phase space volume of $|\Gamma_{M_f}|$, then
$|\Gamma_{M_f}|/|\Gamma_{M_0}| \approx 10^{10^{123}}$.  The high value
of $|\Gamma_{_M{_f}}|$ compared with $|\Gamma_{M_0}|$ comes {}from the
vast amount of phase space corresponding to a universe collapsed into
a black hole, see Fig.\ 2.

I do not know whether these initial and final states are reasonable,
but in any case one has to agree with Feynman's statement [3] that
``it is necessary to add to the physical laws the hypothesis that in
the past the universe was more ordered, in the technical sense, than
it is today...to make an understanding of the irreversibility.''
``Technical sense'' clearly refers to the initial state of the
universe $M_0$ having a smaller $S_B$ than the present state.  Once we
accept such an initial macrostate $M_0$, then the initial microstate
can be assumed to be typical of $\Gamma_{M_0}$.  We can then apply our
statistical reasoning to compute the typical evolution of such an
initial state, i.e. we can use phase-space-volume arguments to predict
the future behavior of macroscopic systems---but not to determine the
past.  As put by Boltzmann [2], ``we do not have to assume a special
type of initial condition in order to give a mechanical proof of the
second law, if we are willing to accept a statistical viewpoint\dots
if the initial state is chosen at random \dots entropy is almost
certain to increase.''
\bigskip
\noindent   {\bf Irreversibility and Macroscopic Stability}

Of course mechanics itself doesn't preclude having a microstate $X$
for which $S_B(M(X_t))$ decreases as $t$ increases.  An
experimentalist could, {\it in principle}, reverse all velocities of
the system in Fig.\ 1b, and then watch the system unmix itself.  It
seems however impossible to do so in practice: Even if he/she managed
to do a perfect job on the velocity reversal part, as occurs
(imperfectly) in spin echo experiments [21] , we would not expect to
see the system in Fig.\ 1 go {}from $M_b$ to $M_a$.  This would
require that {\it both} the velocity reversal and system isolation be
{\it absolutely perfect}.  The reason for requiring such perfection
now and not before is that while the macroscopic behavior of a system
with microstate $Y$ in the state $M_b$ {\it coming} {}from a
microstate $X$ typical with respect to $\Gamma_{M_{\scriptstyle a}}$
is {\it stable} against perturbations as far as its future is
concerned it is very {\it unstable} as far as its {\it past} (and thus
the future behavior of $RY$) is concerned (see Figs.\ 3 and 4).

(I am thinking here primarily of situations like those depicted in
Fig.\ 1 where the macroscopic evolution is described by the stable
diffusion equation. However, even in situations, such as that of
turbulence, where the forward macroscopic evolution is chaotic, i.e.\
sensitive to small perturbations, all evolutions will still have
increasing Boltzmann entropies in the forward direction.  For the
isolated evolution of the velocity reversed microstate, however, one
has decreasing $S_B$ while the perturbed ones can be expected to have,
at least after a very short time, increasing $S_B$.  So even in
macroscopically ``chaotic'' regimes the forward evolution of $M$ is in
this sense much more stable than the backward one. Thus in turbulence
all forward evolutions are still described by solutions of the same
Navier-Stokes equation while the backward macroscopic evolution for a
{\it perfectly isolated} fluid and for an actual one will have no
connection with each other.)

The above analysis is based on the very reasonable assumption that
almost any perturbation of the microstate $Y$ will tend to make it
more typical of its macrostate $M(Y)$, here equal to $M_b$.  The
perturbation will thus not interfere with behavior typical of
$\Gamma_{M_{b}}$. The forward evolution of the unperturbed $RY$ is on
the other hand, by construction, heading towards a smaller phase space
volume and is thus untypical of $\Gamma_{M_{b}}$. It therefore
requires ``perfect aiming'' and will very likely be derailed by even
small imperfections in the reversal and/or tiny outside influences.
After a {\it very short} time in which $S_B$ decreases the
imperfections in the reversal and the ``outside'' perturbations, such
as one coming {}from a sun flare, a star quake in a distant galaxy (a
long time ago) or {}from a butterfly beating its wings [6], will make
it increase again.  This is somewhat analogous to those pinball
machine type puzzles where one is supposed to get a small metal ball
into a particular small region.  You have to do things just right to
get it in but almost anything you do gets it out into larger regions.
For the macroscopic systems we are considering, the disparity between
relative sizes of the comparable regions in the phase space is
unimaginably larger.  In the absence of any ``grand conspiracy'', the
behavior of such systems can therefore be confidently predicted to be
in accordance with the second law (except possibly for very short time
intervals).

Sensitivity to small perturbations in the entropy decreasing direction
is commonly observed in computer simulations of systems with
``realistic'' interactions where velocity reversal is easy to
accomplish but unavoidable roundoff errors play the role of
perturbations. It is possible, however, to avoid this effect in
simulations by the use of discrete time integer arithmetic.  This is
clearly illustrated in Figs.\ 3 and 4.  The latter also shows how a
small perturbation which has no effect on the forward macro evolution
completely destroys the time reversed evolution.  This point is very
clearly formulated in the 1874 paper of Thomson [1]:

``Dissipation of energy, such as that due to heat conduction in a gas,
might be entirely prevented by a suitable arrangement of Maxwell
demons, operating in conformity with the conservation of energy and
momentum.  If no demons are present, the average result of the free
motions of the molecules will be to equalize temperature-differences.
If we allowed this equalization to proceed for a certain time, and
then reversed the motions of all the molecules, we would observe a
disequalization.  However, if the number of molecules is very large,
as it is in a gas, any slight deviation {}from absolute precision in
the reversal will greatly shorten the time during which
disequalization occurs.  In other words, the probability of occurrence
of a distribution of velocities which will lead to disequalization of
temperature for any perceptible length of time is very small.
Furthermore, if we take account of the fact that no physical system
can be completely isolated {}from its surroundings but is in principle
interacting with all other molecules in the universe, and if we
believe that the number of these latter molecules is infinite, then we
may conclude that it is impossible for temperature-differences to
arise spontaneously.  A numerical calculation is given to illustrate
this conclusion.''  Thomson goes on to say: ``The essence of Joule's
discovery is the subjection of physical phenomena to dynamical law.
If, then, the motion of every particle of matter in the universe were
precisely reversed at any instant, the course of nature would be
simply reversed for ever after.  The bursting bubble of foam at the
foot of a waterfall would reunite and descend into the water; \dots
Boulders would recover {}from the mud the materials required to
rebuild them into their previous jagged forms, and would become
reunited to the mountain peak {}from which they had formerly broken
away.  And if also the materialistic hypothesis of life were true,
living creatures would grow backwards, with conscious knowledge of the
future, but no memory of the past, and would become again unborn.  But
the real phenomena of life infinitely transcend human science; and
speculation regarding consequences of their imagined reversal is
utterly unprofitable.  Far otherwise, however, is it in respect to the
reversal of the motions of matter uninfluenced by life, a very
elementary consideration of which leads to the full explanation of the
theory of dissipation of energy.''

\vglue 24pt plus 12pt minus 12pt
\noindent   {\bf  Boltzmann vs. Gibbs Entropies}

The Boltzmannian approach, which focuses on the evolution of a
particular macroscopic system, is conceptually different {}from the
Gibbsian approach, which focuses primarily on ensembles. This
difference shows up strikingly when we compare Boltzmann's
entropy---defined in (1) for a microstate $X$ of a macroscopic
system---with the more commonly used (and misused) entropy $S_G$ of
Gibbs, defined for an ensemble density $\rho(X)$ by
 $$S_G (\{\rho \}) = -k {\textstyle \int} \rho (X) [\log \rho(X)]dX.
\eqno (3)$$

\noindent Here $\rho(X)dX$ is the probability (obtained some way or other)
for the microscopic state of the system to be found in the phase space
volume element $dX$ and the integral is over the phase space $\Gamma$.
Of course if we take $\rho(X)$ to be the generalized microcanonical
ensemble associated with a macrostate $M$,
$$\rho_M(X) \equiv \left \{
\matrix {
| \Gamma_M |^{-1}, & {\rm if}\ X \in \Gamma_M \cr
0, \hfill & {\rm otherwise}\hfill \cr
} \right. ,\eqno (4)\, $$
then clearly,
$$S_G(\{\rho_M \}) = k\log |\Gamma_M | = S_B(M). \eqno (5)$$
Generalized microcanonical ensembles like $\rho_M(X)$, or their
canonical version, are commonly used to describe systems in which the
particle density, energy density and momentum density vary slowly on a
microscopic scale {\it and} the system is, in each small macroscopic
region, in equilibrium with the prescribed local densities, i.e.\ when
we have local equilibrium [10].  In such cases $S_G(\{\rho_M\})$ and
$S_B(M)$ agree with each other, and with the macroscopic hydrodynamic
entropy.

Note however that unless the system is in complete equilibrium and
there is no further systematic change in $M$ or $\rho$, the time
evolutions of $S_B$ and $S_G$ are {\it very} different. As is well
known, it follows {}from the fact that the volume of phase space
regions remains unchanged under the Hamiltonian time evolution (even
though their shape changes greatly) that $S_G(\{ \rho \})$ never
changes in time as long as $X$ evolves according to the Hamiltonian
evolution, i.e.\ $\rho$ evolves according to the Liouville equation;
$S_B(M)$, on the other hand, certainly does change. Thus, if we
consider the evolution of the microcanonical ensemble corresponding to
the macrostate $M_a$ in Fig.\ 1a after removal of the constraint,
$S_G$ would equal $S_B$ initially but subsequently $S_B$ would
increase while $S_G$ would remain constant. $S_G$ therefore does not
give any indication that the system is evolving towards equilibrium.

This reflects the fact, discussed earlier, that the microstate
 $X(t)$ does not remain typical of the local equilibrium state
$M(t)$ for $t > 0$. As long as the system remains truly isolated the state
$T_tX$ will contain subtle correlations, which are reflected in the
complicated shape which an initial region $\Gamma_M$ takes on in time
but
which do not affect the future time evolution of $M$ (see the discussion at end
of section on Initial Conditions).  {\it Thus the relevant
entropy for understanding the time evolution of macroscopic systems is $S_B$
and not $S_G$}.  (Of course if we are willing to do a ``course graining''
of $\rho$ over cells $\Gamma_M$ then we are essentially back to dealing
with $\rho_M$, or  superpositions of such  $\rho_M$'s and we are
just defining $S_B$ in a backhanded way.)
\bigskip
\noindent {\bf Remarks}

a) The characterization of a macrostate $M$ usually done via density
fields in three dimensional space as in Fig.\ 1 can be extended to
mesoscopic descriptions.  This is particularly convenient for a {\it
dilute gas} where $M$ can be usefully characterized by the density in
the six dimensional position and velocity space of a single molecule.
The deterministic macroscopic (or mesoscopic) evolution of this $M$ is
{\it then} given by the Boltzmann equation and $S_B(M)$ coincides with
the negative of Boltzmann's famous $H$-function.

It is important to note however that for systems in which the
potential energy is relevant, e.g.\ non-dilute gases, the $H$-function
does not agree with $S_B$ and $-H$ (but not $S_B$) will decrease for
suitable macroscopic initial conditions. As pointed out by Jaynes [24]
this will happen whenever one starts with an initial total energy $E$
and kinetic energy $K=K_0$ such that $K_0 > K_{eq} (E)$, the value
that $K$ takes when the system is in equilibrium with energy $E$.
This can be readily seen if the initial macrostate is one in which the
spatial density is uniform and the velocity distribution is Maxwellian
with the appropriate temperature $T_0 = {2 \over 3} K_0/kN$.  The
temperature will then decrease as the system goes to equilibrium and
$-H$ which, for a Maxwellian distribution, is proportional to $\log T$
will therefore be smaller in the equilibrium state when $T = T_{eq}
(E) < T_0$.

\noindent b)  Einstein's 
formula for the probability of fluctuations in an equilibrium system,
$${\rm Probability ~~ of ~~} M ~~~ \sim ~~~ \exp\{[S(M) -
S_{eq}]/k\}$$ is essentially an inversion of formulas (4) and (5).
When combined with the observation that the entropy $S_B(M)$ of a
macroscopic system, prepared in a specified nonuniform state $M$, can
be computed {}from macroscopic thermodynamic considerations it yields
useful results.  In particular when $S_B(M)$ in the exponent is
expanded around $M_{eq}$, and only quadratic terms are kept, we obtain
a Gaussian distribution for normal (small) fluctuations {}from
equilibrium.  This is one of the main ingredients of Onsager's
reciprocity relations [25].

\bigskip
\noindent   {\bf Typical vs. Averaged Behavior}

I conclude by emphasizing again that having results for typical
microstates rather than averages is not just a mathematical nicety but
goes to the heart of the problem of understanding the microscopic
origin of observed macroscopic behavior --- {\it we neither have nor
do we need ensembles when we carry out observations like those
illustrated in Fig.\ 1.} What we do need and can expect to have is
typical behavior.  Ensembles are merely mathematical tools, useful as
long as the dispersion, in the quantities we are interested in, is
sufficiently small.  This is always the case for properly defined
macroscopic variables in equilibrium Gibbs ensembles. The use of such
an ensemble as the initial ``statistical state" immediately following
the lifting of a constraint {}from a macroscopic system in equilibrium
at some time $t_0$ is also sensible, as long as the evolution of
$M(t)$ is, with probability close to one, the same for all systems in
the ensemble.

There is no such typicality with respect to ensembles describing the
time evolution of a system with only a few degrees of freedom. This is
an essential difference (unfortunately frequently overlooked or
misunderstood) between the irreversible and the chaotic behavior of
Hamiltonian systems.  The latter, which can be observed already in
systems consisting of only a few particles, will not have a
uni-directional time behavior in any particular realization.  Thus if
we had only a few hard spheres in the box of Fig.\ 1, we would get
plenty of chaotic dynamics and very good ergodic behavior (mixing,
K-system, Bernoulli) but we could not tell the time order of any
sequence of snapshots.

Finally I note that my discussion has focused exclusively on what is
usually referred to as the thermodynamic arrow of time and on its
connection with the cosmological initial state.  I did not discuss
other arrows of time such as the asymmetry between advanced and
retarded electromagnetic potentials or ``causality''.  It is my
general feeling that these are all manifestations of the low entropy
initial state of the universe.  I also believe that the violation of
time reversal invariance in the weak interactions is not relevant for
macroscopic irreversibility.

\bigskip
\noindent {\bf Acknowledgments:} I want to thank Y. ~Aharonov, G.~Eyink, O.
~Penrose and especially S.~Goldstein, H.~Spohn and E.~Speer for many
very useful discussions.  I also thank the organizers of this
conference for their kind hospitality.  Various aspects of this
research have been supported in part by the AFOSR and NSF.

\bigskip

\noindent {\bf References}
\bigskip
\openup-2\jot
\parskip=6pt
\item {[1]} \ W. Thomson, Proc. of the Royal Soc. of Edinburgh, {\bf 8} 325
(1874), reprinted in 1a).

\item {[2]}  \ a)  For a collection of original articles of Boltzmann
and others 
{}from the second half of the nineteenth century on this subject (all
in English) see S.G. Brush, {\it Kinetic Theory}, Pergamon, Oxford,
(1966).
\item\item {b)} \ For an interesting biography of Boltzmann, which also
contains many references, see E. Broda {\it Ludwig Boltzmann,
Man---Physicist---Philosopher}, Ox Bow Press, Woodbridge, Conn (1983); 
translated {}from the German.
\item \item {c)} For a historical discussion of Boltzmann and his
ideas see also articles by M. Klein, E. Broda, L. Flamn 
in {\it The Boltzmann Equation, Theory and Application}, E.G.D. Cohen
and W. Thirring, eds., Springer-Verlag, 1973.
\item \item {d)} \ For a general history of the subject see S.G. Brush,
{\it The
Kind of Motion We Call Heat}, Studies in Statistical Mechanics, vol.
VI, E.W. Montroll and J.L. Lebowitz, eds.  North-Holland, Amsterdam,
(1976).
\item \item {e)}  \ G. Gallavotti, Ergodicity, Ensembles,
Irreversibility in Boltzmann and Beyond, J. Stat. Phys., to
appear, (1995).
\item {[3]} \ R. Feynman, {\it The Character of Physical Law}, MIT P.,
Cambridge, Mass. (1967), ch.5. R.P Feynman, R. B Leighton, M. Sands, {\it
The Feynman Lectures on Physics}, Addison-Wesley, Reading, Mass. (1963),
sections 46--3, 4, 5.
\item {[4]} \ O. Penrose, {\it Foundations of Statistical Mechanics},
Pergamon, Elmsford, N.Y. (1970), ch. 5.
\item {[5]} \ R. Penrose, {\it The Emperor's New Mind}, Oxford U. P.,
New York (1990), ch. 7.
\item {[6]} \ D. Ruelle, {\it Chance and Chaos}, Princeton U. P.,
Princeton, N.J. (1991), ch. 17, 18.
\item {[7]} \ O. Lanford, Physica A {\bf 106}, 70 (1981).
\item {[8]} \ J.L. Lebowitz, Physica A {\bf 194}, 1 (1993).
\item {[9]} \ J.C. Maxwell, {\it Theory of Heat}, p. 308:  ``Tait's
Thermodynamics'', Nature {\bf 17}, 257 (1878).  Quoted in M. Klein, ref.
1c).
\item {[10]} \ H. Spohn, {\it Large Scale Dynamics of Interacting
Particles}, Springer-Verlag, New York (1991).  A. De Masi, E. Presutti,
{\it Mathematical Methods for Hydrodynamic Limits}, Lecture Notes in Math
1501, Springer-Verlag, New York (1991).  J.L. Lebowitz, E. Presutti, H.
Spohn, J. Stat. Phys. {\bf 51}, 841 (1988).
\item {[11]} \ Y. Aharonov, P.G. Bergmann, J.L. Lebowitz, Phys. Rev. B {\bf
134}, 1410 (1964).  D.N. Page, Phys. Rev. Lett. {\bf 70}, 4034 (1993).
\item {[12]} \ J.S. Bell, {\it Speakable and Unspeakable in Quantum
Mechanics}, Cambridge U. P., New York (1987).
\item {[13]} \ D. D{\"u}rr, S. Goldstein, N. Zanghi, J. Stat. Phys. {\bf
67}, 843 (1992).  
\item {[14]} \ See articles by M. Gell-Mann, J. Hartle, R. Griffiths,
D. Page and 
others in {\it Physical Origin of Time Asymmetry}, J.J. Halliwell,
J. Perez-Mercader and W.H.  Zurek, eds., Cambridge University Press,
1994.
\item {[15]} \ {\it Vorlesungen {\"u}ber Gastheorie}.  2 vols.
Leipzig:  Barth, 
1896, 1898.  This book has been translated into English by S.G. Brush, {\it
Lectures on Gas Theory}, (London:  Cambridge University Press, 1964)
.
\item {[16]} \ J.W. Gibbs, Connecticut Academy Transactions {\bf 3}, 229
(1875), reprinted in {\it The Scientific Papers}, {\bf 1}, 167 (New
York, 1961).
\item {[17]} \ L. Boltzmann, Ann. der Physik {\bf 57}, 773 (1896).
Reprinted in 1a).
\item {[18]} \ E. Schr{\"o}dinger, {\it What is Life?  And Other Scientific
Essays}, Doubleday Anchor Books, New York (1965), section 6.
\item {[19]} \ J.L. Lebowitz and H. Spohn, {\it Communications on Pure
and Applied  Mathematics}, 
XXXVI,595, (1983); see in particular section 6(i).
\item {[20]} \ L. Boltzmann (1886) quoted in E. Broda, 1b), p. 79.
\item {[21]} \ E.L. Hahn, Phys. Rev. {\bf 80}, 580 (1950).  See  also S.
Zhang, B.H. Meier, R.R. Ernst, Phys. Rev. Let.. {\bf 69} 2149 (1992).
\item {[22]} \ D. Levesque and L. Verlet, J. Stat.
Phys. {\bf 72}, 519 (1993).
\item {[23]} \ B.T. Nadiga, J.E. Broadwell and B. Sturtevant, {\it
Rarefield Gas 
Dynamics:  Theoretical and Computational Techniques}, edited by E.P. Muntz,
D.P. Weaver and D.H. Campbell, Vol 118 of {\it Progress in Astronautics 
and Aeronautics}, AIAA, Washington, DC, ISBN 0--930403--55--X, 1989.
\item {[24]} \ E.T. Jaynes, Phys. Rev. A{\bf 4}, 747 (1971).
\item {[25]} \ A. Einstein, Am. Phys. (Leipzig) {\bf 22}, 180 (1907); {\bf
33}, 1275 (1910).  L. Onsager, Phys. Rev. {\bf 37}, 405 (1931); {\bf 38},
2265 (1931).
\bigskip
\noindent {\bf Figure Captions}
\bigskip
\noindent {\bf Fig.\ 1}  How would you order this sequence of ``snapshots''
in time?  Each represents a macroscopic state of a system containing, for
example, two differently colored fluids.
\bigskip
\noindent {\bf Fig.\ 2}  With a gas in a box, the maximum entropy state
(thermal equilibrium) has the gas distributed uniformly; however, with a
system of gravitating bodies, entropy can be increased {}from the uniform
state by gravitational clumping leading eventually to a black hole.
{}From Ref.\ [5]. 
\bigskip
\noindent {\bf Fig.\ 3}  Time evolution of a system of 900 particles all  
interacting
via the same cutoff Lennard-Jones pair potential using integer
arithmetic.  Half of the particles are colored white, the other half black.
All velocities are reversed at $t=20,000$.  The system then retraces its
path and the initial state is fully recovered.  {}From Ref.\ [22].
\bigskip
\noindent {\bf Fig.\ 4}  Time evolution of a reversible cellular automaton
lattice gas using integer arithmetic.   Figures a) and c) show the
mean velocity,  figures b) and d) 
the entropy.  The mean velocity 
decays with time and the entropy increases up to $t=600$ when there is a 
reversal of all  velocities.  The system then retraces its path and
the initial state is 
fully recovered in figures a) and b).  In the bottom figures there is
a small error in the 
reversal at $t=600$.  While such an error has no appreciable effect
on the initial evaluation it effectively prevents any recovery of the
macroscopic velocity.  The entropy, on the scale of the figure, just
remains at its maximum value.  This shows
the instability of the reversed path.  {}From Ref.\ [23].
\end